\documentclass[runningheads]{llncs}
\usepackage{graphicx}
\usepackage{booktabs}
\usepackage{caption}
\usepackage{rotating}
\usepackage{multirow}
\usepackage{booktabs}
\usepackage{adjustbox}
\usepackage{pifont}
\usepackage{amsmath}
\usepackage{colortbl} 
\definecolor{lightyellow}{RGB}{255, 255, 204} 
\definecolor{lightgreen}{RGB}{204, 255, 204} 
\definecolor{lightgray}{RGB}{220, 220, 220}    
\usepackage{xcolor}   
\usepackage{graphicx} 
\usepackage{multirow} 
\usepackage{makecell} 

\usepackage{graphicx} 
\usepackage{transparent} 
\begin{document}
\vspace{-5cm}   
\title{MDAA-Diff: CT-Guided Multi-Dose Adaptive Attention Diffusion Model for PET Denoising}
\titlerunning{MDAA-Diff}
\author{Xiaolong Niu$^{1}$, Zanting Ye$^{1}$, Xu Han$^{2}$, Yanchao Huang$^{3}$, Hao Sun$^{1}$, Hubing Wu$^{3}$, Lijun Lu$^{1}$*}  
\authorrunning{Niu et al.}
\institute{$^{1}$School of Biomedical Engineering, Southern Medical University, Guangzhou, China \\
\email{ljlubme@gmail.com}\\
$^{2}$School of Biomedical Engineering, Shanghai Jiao Tong University,shanghai, China \\
$^{3}$Nanfang PET Center, Nanfang Hospital Southern Medical University, Guangzhou, China \\}

%
%
\maketitle        
\begin{abstract}
Acquiring high-quality Positron Emission Tomography (PET) images requires administering high-dose radiotracers, which increases radiation exposure risks. Generating standard-dose PET (SPET) from low-dose PET (LPET) has become a potential solution. However, previous studies have primarily focused on single low-dose PET denoising, neglecting two critical factors: discrepancies in dose response caused by inter-patient variability, and complementary anatomical constraints derived from CT images. In this work, we propose a novel CT-Guided Multi-dose Adaptive Attention Denoising Diffusion Model (MDAA-Diff) for multi-dose PET denoising. Our approach integrates anatomical guidance and dose-level adaptation to achieve superior denoising performance under low-dose conditions. Specifically, this approach incorporates a CT-Guided High-frequency Wavelet Attention (HWA) module, which uses wavelet transforms to separate high-frequency anatomical boundary features from CT images. These extracted features are then incorporated into PET imaging through an adaptive weighted fusion mechanism to enhance edge details. Additionally, we propose the Dose-Adaptive Attention (DAA) module, a dose-conditioned enhancement mechanism that dynamically integrates dose levels into channel-spatial attention weight calculation. Extensive experiments on $^{18}$F-FDG and $^{68}$Ga-FAPI datasets demonstrate that MDAA-Diff outperforms state-of-the-art approaches in preserving diagnostic quality under reduced-dose conditions. Our code is publicly available.

\end{abstract}
\section{Introduction}
Positron Emission Tomography (PET) is a molecular imaging technique used to evaluate metabolic activity and functional states of human tissues and organs. PET facilitates precise functional assessments of tissues and organs through tracer-based metabolic visualization imaging, demonstrating indispensable clinical significance in tumor diagnosis, neurodegenerative disease research, and cardiovascular functional analysis\cite{ref2,ref8,ref27}. However, PET imaging quality remains highly dependent on radiotracer dose. While low-dose PET (LPET) imaging mitigates the risk of ionizing radiation (particularly for sensitive populations such as children and pregnant women~\cite{ref3,ref4,ref11}), it inevitably increases image noise, compromising diagnostic accuracy. This inherent trade-off between radiation reduction and diagnostic efficacy remains a major challenge in contemporary PET imaging technology.

Deep learning has emerged as a pivotal approach for radiation dose reduction in PET imaging, enabling diagnostic-quality standard-dose PET (SPET) denoising from LPET. Compared with traditional methods, such as local mean filtering and frequency-domain filtering~\cite{ref1,ref5,ref22}, which often oversmooth critical details, deep neural networks learn complex nonlinear mappings between LPET and SPET through end-to-end training~\cite{ref6,ref9,ref19,ref21,ref23,ref24,ref25}, suppressing noise while preserving lesion metabolic activity features. Generative model-based denoising architectures have become a focal point. For instance, Sanaat et al.~\cite{ref16} utilized CycleGAN to achieve cross-domain mapping from LPET to SPET. Zhou et al.~\cite{ref26} combined StyleGAN with segmentation network to generate denoised images with realistic textures. Nevertheless, these deterministic mapping models struggle to quantify uncertainties inherent in the reconstruction process. Denoising Diffusion Probabilistic Model (DDPM)~\cite{ref10} addresses this limitation by iteratively denoising images through a Markov chain, enabling uncertainty modeling while retaining fine-grained details. However, existing approaches predominantly assume single LPET denoising, ignoring the variations in dose response caused by inter-patient variability in clinical practice, such as physiology, metabolism, and tracer uptake patterns.

Preliminary progress has been achieved in research for multi-dose PET denoising. Xie et al.~\cite{ref7} proposed a Unified Noise-aware Network that performs PET denoising across different dose levels. But its weight allocation mechanism compromises recovery performance under high-dose conditions. Tang et al.~\cite{ref18} developed a High-Frequency-Guided Residual Diffusion Model which uses sinusoidal positional encoding to encode the dose, and then input it into the denoising network along with the embedded timesteps. However, the simple concatenation of dose embedding and time embedding may fail to enable effective interaction between dose information and the diffusion process, potentially limiting the model's adaptability to feature variations across different dose levels. Moreover, this method inadequately exploits the advantages of multi-modal imaging in clinical applications. In PET/CT multi-modal imaging, the high spatial coherence between CT and PET images provides critical anatomical constraints that enhance denoising performance~\cite{ref7}.

In this work, we propose a CT-Guided Multi-Dose Adaptive Attention Diffusion Model (MDAA-Diff) for PET denoising. Specifically, we design a CT-Guided High-frequency Wavelet Attention (HWA) module to extract CT high-frequency anatomical boundary features using multi-scale wavelet decomposition, which are then adaptively fused with PET imaging. Moreover, we construct a Dose-adaptive Attention (DAA) module that utilizes dose factors to guide channel-wise and spatial feature weighting, enhancing the model's perception capability and robustness under different dose conditions.

\begin{figure*}[!t]
\centerline{\includegraphics[width=0.95\linewidth]{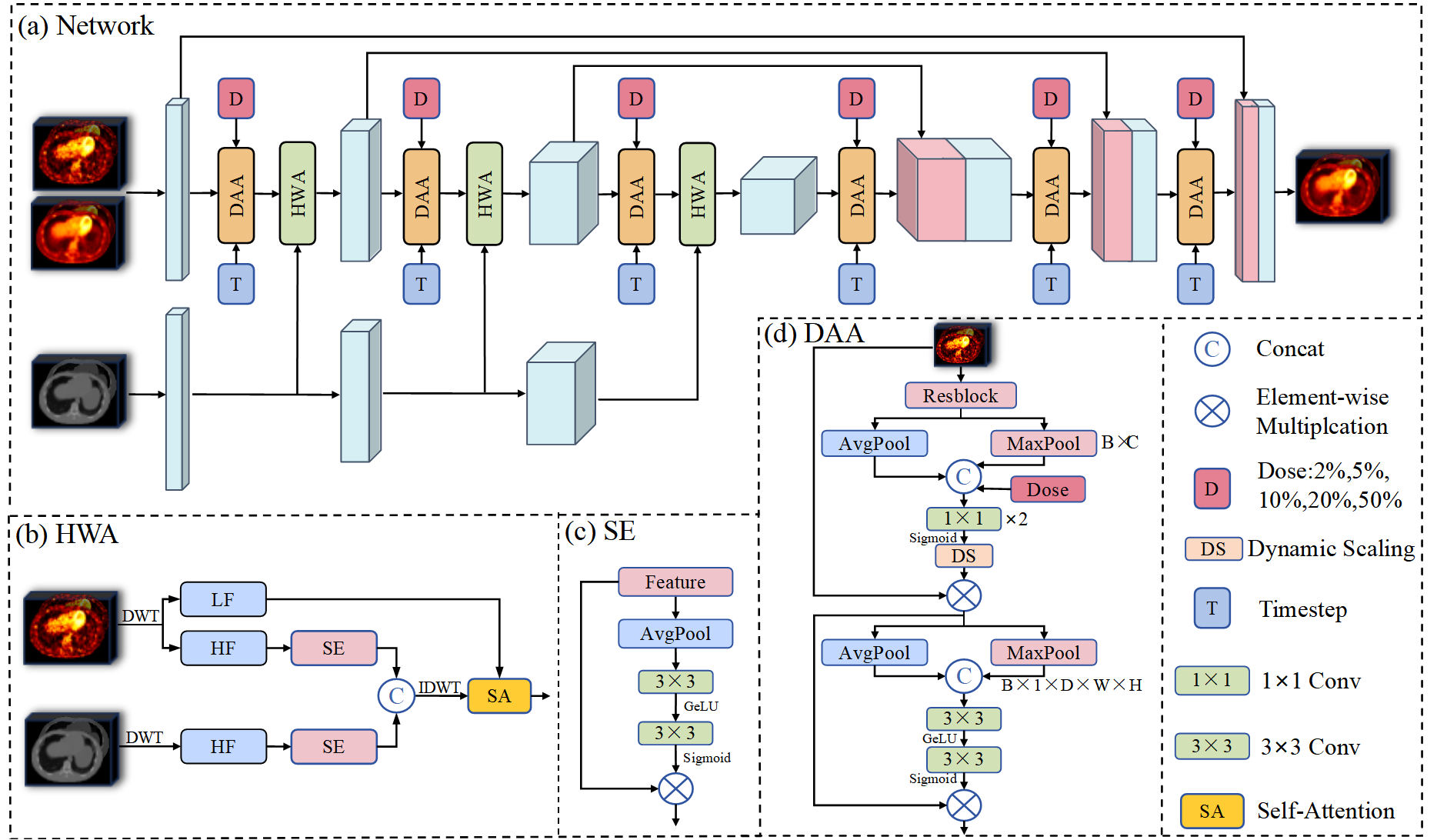}}
\caption{Overview of the MDAA-Diff. (a) is the overall network architecture of the diffusion model; (b) is a CT-guided modal fusion module; (c) is designed to perform channel-wise attention; and (d) is a multi-dose adaptive attention module.}
\label{fig:flow}
\end{figure*}

\section{Methodology}
\subsection{Improved Denoising Diffusion Probabilistic Models}
Our model builds upon an Improved Denoising Diffusion Probabilistic Models (IDDPM)~\cite{ref14}, which significantly reduces the sampling timesteps compared to DDPM. While DDPM models the reverse denoising process as a conditional probability distribution, defined as:
\begin{equation}
p_\theta(x_{t-1}|x_t) = \mathcal{N}(x_{t-1};\, \mu_\theta(x_t, t),\, \Sigma_\theta(x_t, t)),
\end{equation}
where $\mu_\theta(x_t, t)$ represents the learnable value, and the variance $\Sigma_\theta(x_t, t)$ is fixed as $\sigma^2 I$ and cannot be optimized as a learnable parameter. IDDPM addresses this limitation by parameterizing $\Sigma_\theta(x_t, t)$ modeling through the following formulation:
\begin{equation}
\Sigma_\theta(x_t, t) = \exp \big( v \log \beta_t + (1-v) \log \tilde{\beta}_t \big),
\end{equation}
where $v$ is a weighting parameter, $\beta_t$ represents the noise intensity during the forward diffusion process, and $\tilde{\beta}_t$ denotes the adjusted noise intensity in the reverse process. This parameterization enhances both the model’s flexibility and the quality of the generated samples. Additionally, it optimizes noise scheduling by selecting specific timesteps $S = \{s_1, s_2, \ldots, s_K\}$, which further improves sampling efficiency.

\subsection{MDAA-Diff Model}
Our MDAA-Diff model is shown in Fig.~\ref{fig:flow}. First, the noise-perturbed SPET and LPET images are concatenated along the channel dimension and fed into the input model. Multi-scale PET/CT features are extracted by a dedicated encoder to achieve cross-modal information fusion. The dose factor is dynamically integrated into PET feature enhancement through the DAA module, enabling dose-adaptive adjustments.
\subsubsection{High-frequency Wavelet Attention}
Current PET and CT feature fusion methods often rely on channel concatenation or Cross-Attention (CA) mechanisms~\cite{ref13}. The quadratic-complexity operation of CA imposes prohibitive computational demands for high-resolution medical imaging. To address this issue, we propose the HWA module, a fusion module based on 3D wavelet transform and Squeeze-and-Excitation (SE), to achieve multi-scale cooperative enhancement of cross-modal features. As shown in Fig.~1b, PET and CT features are decomposed into low-frequency and high-frequency components: PET low-frequency features capture global metabolic patterns, while high-frequency features encode lesion edges. CT features are dominated by high-frequency components, which contain high-resolution anatomical information such as bone boundaries and organ morphology, and provide accurate spatial localization and structural constraints for the PET images.

On this basis, the SE module (as shown in Fig.~1c) is utilized to perform dynamic channel weights for both PET and CT high-frequency components, emphasizing key features and reducing redundancy. The weighted PET and CT high-frequency features are concatenated along the channel dimension and processed through convolutional layers for cross-modal interaction and dimensionality reduction. To enable multi-scale information complementarity, HWA modules are embedded across all encoder levels (from shallow local details to deep semantic features) to establish a progressive fusion architecture. Finally, the fused features are reconstructed from the PET low-frequency components and the convolved high-frequency components through inverse wavelet transform.

\subsubsection{Dose-Adaptive Attention Module}
The DAA module adaptively enhances multi-dose features through the synergistic mechanism of the channel enhancement branch and the spatial enhancement branch. The details of the HWA module are shown in Fig.~1d. The module dynamically adjusts the weights of the input features to accommodate different dose-level distributions and improve the robustness of the model.

\textit{The channel feature enhancement branch.} The branch primarily focuses on dynamically adjusting the channel weights of the input features by capturing global and local information, while integrating the dose factor $D$ to achieve adaptive optimization. Specifically, $D$ is projected through a multi-layer perceptron (MLP) to generate the dose embedding. Global average pooling and max pooling operations are applied to the input features, and the results are concatenated with the dose embedding. The channel weights are then generated through a convolutional network, defined as:
\begin{equation}
W_{\text{channel}} = \mathrm{sigmoid}\Big(W_2 \cdot \mathrm{ReLU}\big(W_1 \cdot \mathrm{Concat}(\mathrm{Pool}(X), D_{\mathrm{emb}})\big)\Big)
\end{equation}
where $\mathrm{Pool}(X) \in \{\mathrm{AvgPool}(X),\, \mathrm{MaxPool}(X)\}$, $W_1$ and $W_2$ are $1 \times 1$ convolutional weight matrices. The channel weights are further scaled by the factor $\alpha$ and shifted by the bias factor $\beta$, which are used to normalize and adjust the input features, achieving dynamic channel enhancement, defined as:
\begin{equation}
X' = \mathrm{Norm}(X) \odot (1 + \alpha) + \beta
\label{eq:4}
\end{equation}
where $\mathrm{Norm}(X)$ represents a normalization operation, and $\odot$ denotes element-wise multiplication.

\textit{The spatial feature enhancement branch.} The branch aims to extract the structural detail and local texture information of the input features along the spatial dimension. Specifically, the input features are pooled along the channel dimension using global average pooling and max pooling. The results are concatenated and passed through a convolutional layer to generate the spatial attention map, defined as:
\begin{equation}
X' = X \odot \mathrm{sigmoid}\left( \mathrm{Conv}\left( \mathrm{Concat}(\mathrm{Mean}(X), \mathrm{Max}(X)) \right) \right),
\label{eq:5}
\end{equation}
where $\mathrm{Conv}$ is a convolutional block composed of a $3 \times 3$ kernel and a dilated convolution (dilation rate set to 2). Finally, the spatial attention map is multiplied element-wise with the input features to achieve context enhancement along the spatial dimension.

\section{Experiments}
\subsection{Datasets}
We evaluated the performance of MDAA-Diff using two independent PET/CT datasets comprising 51 patients with $^{18}$F-FDG tracer and 60 patients with $^{68}$Ga-FAPI tracer. SPET images were reconstructed from 300 seconds (100\%) acquisitions obtained 60 minutes post-injection. LPET images were reconstructed based on time windows resampled during acquisition, simulating reduced acquisition durations of 6 seconds (2\%), 15 seconds (5\%), 30 seconds (10\%), 60 seconds (20\%), and 150 seconds (50\%). All PET images were reconstructed with resolution of $192 \times 192 \times 673$. To address PET-CT mismatches, CT images were resampled and registered to the PET coordinate system using mutual information optimization.

\subsection{Implementation Details}
Our MDAA-Diff model was implemented in PyTorch and trained on an NVIDIA GeForce RTX 4090 GPU with 24GB of memory. A five-fold cross-validation strategy was employed in the experimental design. The number of diffusion steps during the training phase was set to 1000, and the number of sampling steps during the inference phase was set to 50. We used an AdamW optimizer with an initial learning rate of 1e-4, utilizing the cosine annealing scheduling strategy. The generated image quality was evaluated using the Peak Signal-to-Noise Ratio (PSNR) and Structural Similarity Index (SSIM) metrics. Visual comparisons were conducted to demonstrate the model’s ability to improve anatomical fidelity and effectively suppress noise in critical regions.


\begin{figure*}[t]
\centering
\includegraphics[width=0.9\textwidth]{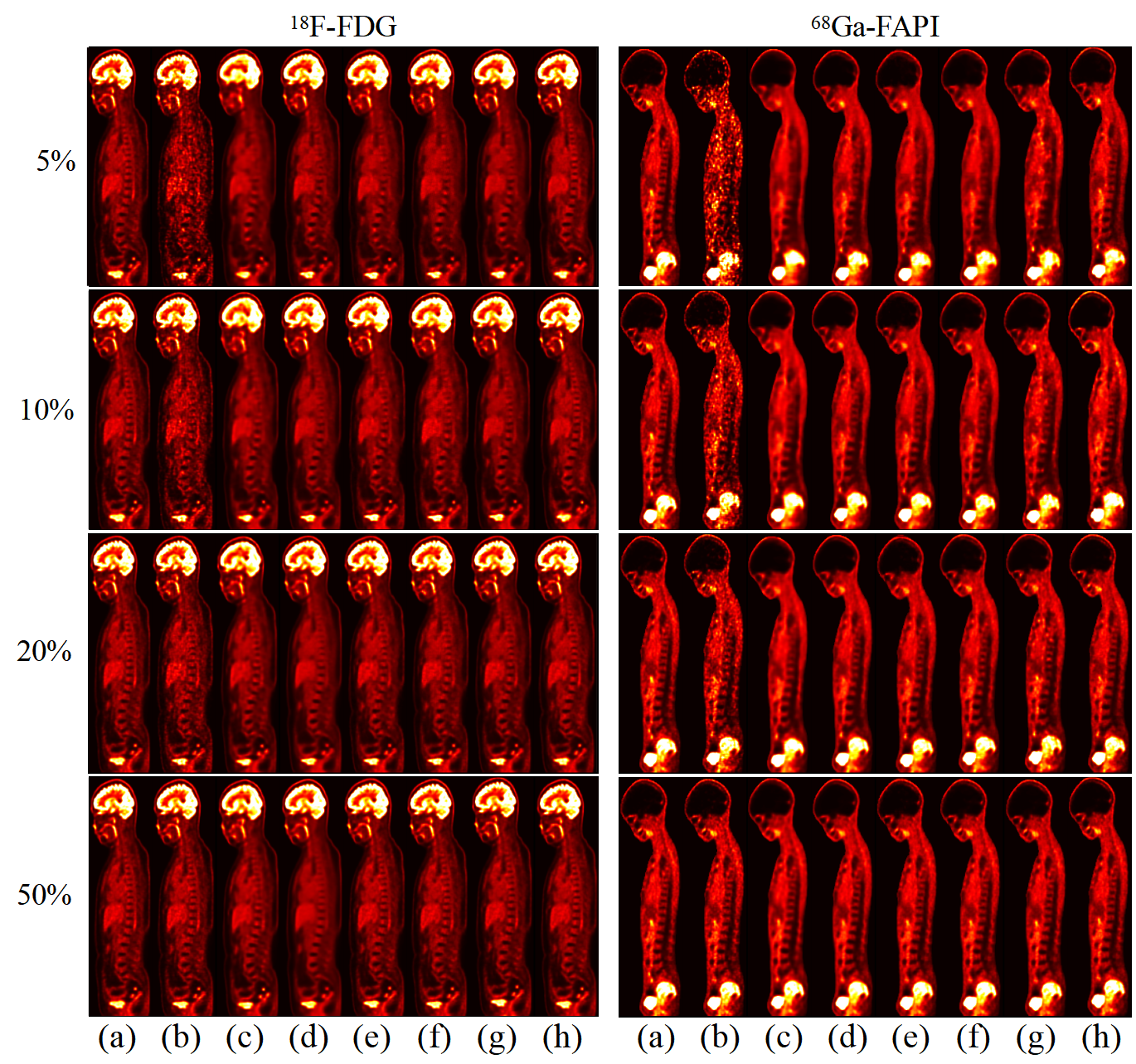} 
\caption{Overview of the MDAA-Diff. (a) is the overall network architecture of the diffusion model; (b) is a CT-guided modal fusion module; (c) is designed to perform channel-wise attention; and (d) is a multi-dose adaptive attention module.}
\label{fig:Ablation}
\vspace{-0.5cm}  
\end{figure*}

\begin{table}[h]
    \centering
    \caption{Quantitative comparison of MDAA-Diff with five state-of-the-art methods. PSNR and SSIM are reported for both tracers at different dose levels.}
    \label{tab:comparison}
    \setlength{\tabcolsep}{8pt}
    \resizebox{\textwidth}{!}{
    \begin{tabular}{@{}lccccc@{}}
        \toprule
        \multicolumn{6}{c}{$^{18}$F-FDG} \\
        \cmidrule(lr){1-6}
        PSNR$\uparrow$/SSIM$\uparrow$ & 6s & 15s & 30s & 60s & 150s \\
        \midrule
        CNN~\cite{ref17}        & 39.858/0.969 & 40.082/0.972 & 40.269/0.974 & 40.459/0.977 & 40.614/0.978 \\
        SwinUnetr~\cite{ref15}   & 44.408/0.976 & 44.979/0.979 & 45.417/0.981 & 46.059/0.983 & 47.115/0.986 \\
        MambaMIR~\cite{ref12}    & 44.349/0.978 & 45.801/0.982 & 45.830/0.983 & 46.043/0.985 & 46.774/0.987 \\
        UNN~\cite{ref20}         & 44.504/0.972 & 45.185/0.977 & 45.663/0.980 & 46.071/0.982 & 46.646/0.984 \\
        HF-ResDiff~\cite{ref18}  & 44.728/0.979 & 45.243/0.982 & 45.677/0.984 & 46.174/0.986 & 47.004/0.984 \\
        \textbf{Ours}            & \textbf{46.035/0.984} & \textbf{46.523/0.987} & \textbf{46.889/0.988} & \textbf{47.315/0.989} & \textbf{48.042/0.990} \\
        \midrule
        \multicolumn{6}{c}{$^{68}$Ga-FAPI} \\
        \cmidrule(lr){1-6}
        PSNR$\uparrow$/SSIM$\uparrow$ & 6s & 15s & 30s & 60s & 150s \\
        \midrule
        CNN~\cite{ref17}        & 39.570/0.967 & 39.832/0.961 & 40.049/0.963 & 40.276/0.965 & 40.680/0.968 \\
        SwinUnetr~\cite{ref15}   & 44.119/0.966 & 44.560/0.969 & 44.866/0.971 & 45.171/0.973 & 45.783/0.976 \\
        MambaMIR~\cite{ref12}    & 44.028/0.968 & 44.429/0.970 & 44.810/0.972 & 45.207/0.974 & 45.844/0.977 \\
        UNN~\cite{ref20}         & 44.759/0.969 & 45.019/0.973 & 45.432/0.975 & 45.765/0.977 & 46.496/0.980 \\
        HF-ResDiff~\cite{ref18}  & 44.219/0.964 & 44.632/0.968 & 45.081/0.971 & 45.493/0.974 & 46.287/0.977 \\
        \textbf{Ours}            & \textbf{46.689/0.980} & \textbf{47.009/0.981} & \textbf{47.160/0.981} & \textbf{47.362/0.982} & \textbf{47.921/0.983} \\
        \bottomrule
    \end{tabular}
    }
\end{table}

\subsection{Comparison with State-of-the-Art Methods}
We compared MDAA-Diff with five state-of-the-art methods, including single-dose approaches (CNN~\cite{ref17}, SwinUnetr~\cite{ref15}, MambaMIR~\cite{ref12}) and multi-dose approaches (HF-ResDiff~\cite{ref18}, UNN~\cite{ref20}). Quantitative analysis (as shown in Table~\ref{tab:comparison}) demonstrated that MDAA-Diff outperformed all comparison models across both PSNR and SSIM metrics. Moreover, its performance improved significantly at lower dose levels (e.g., 2\% and 5\% doses). For instance, in 2\% dose imaging for the $^{18}$F-FDG tracer, MDAA-Diff achieved an SSIM of 98.4\%, which is 0.5 percentage points higher than the second-best model (HF-ResDiff: 97.9\%). This finding highlights its robustness to complex noise distributions.

Qualitative analysis further validated the clinical value of the proposed method. As illustrated in Fig.~\ref{fig:Ablation}, MDAA-Diff exhibited significant improvements in boundary sharpness for the spine region. Particularly in the 2\% dose experiments, its structural fidelity was noticeably superior to that of other comparison methods. For denoising tasks of small lesions, MDAA-Diff effectively suppressed noise artifacts (as indicated by the blue arrows in Fig.~\ref{fig:compare}) while preserving the fine texture details of the lesions.

\begin{figure*}[!t]
\centerline{\includegraphics[width=0.9\linewidth]{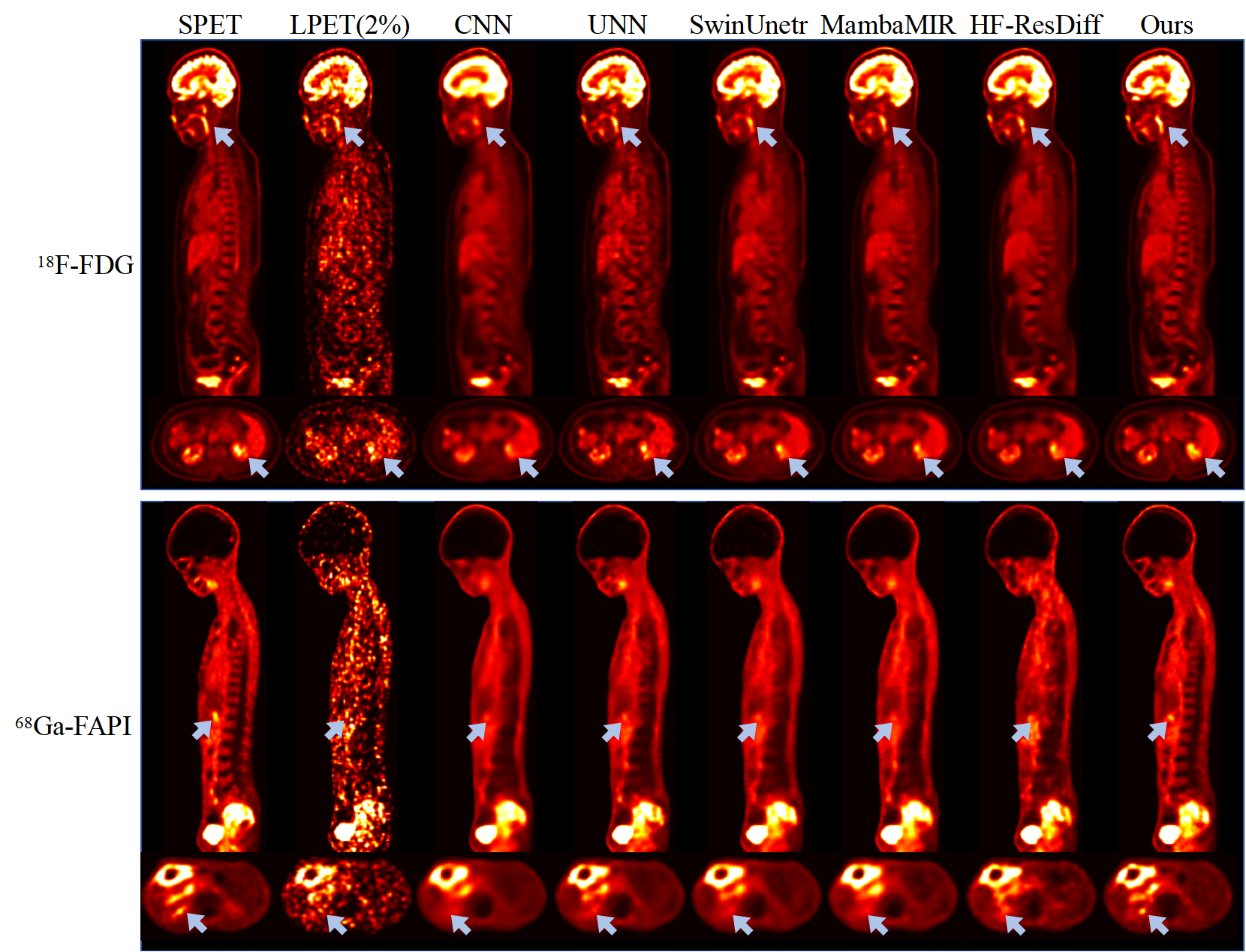}}
\caption{Overview of the MDAA-Diff. (a) is the overall network architecture of the diffusion model; (b) is a CT-guided modal fusion module; (c) is designed to perform channel-wise attention; and (d) is a multi-dose adaptive attention module.}
\label{fig:compare}
\end{figure*}

\begin{figure*}[!t]
\centerline{\includegraphics[width=0.9\linewidth]{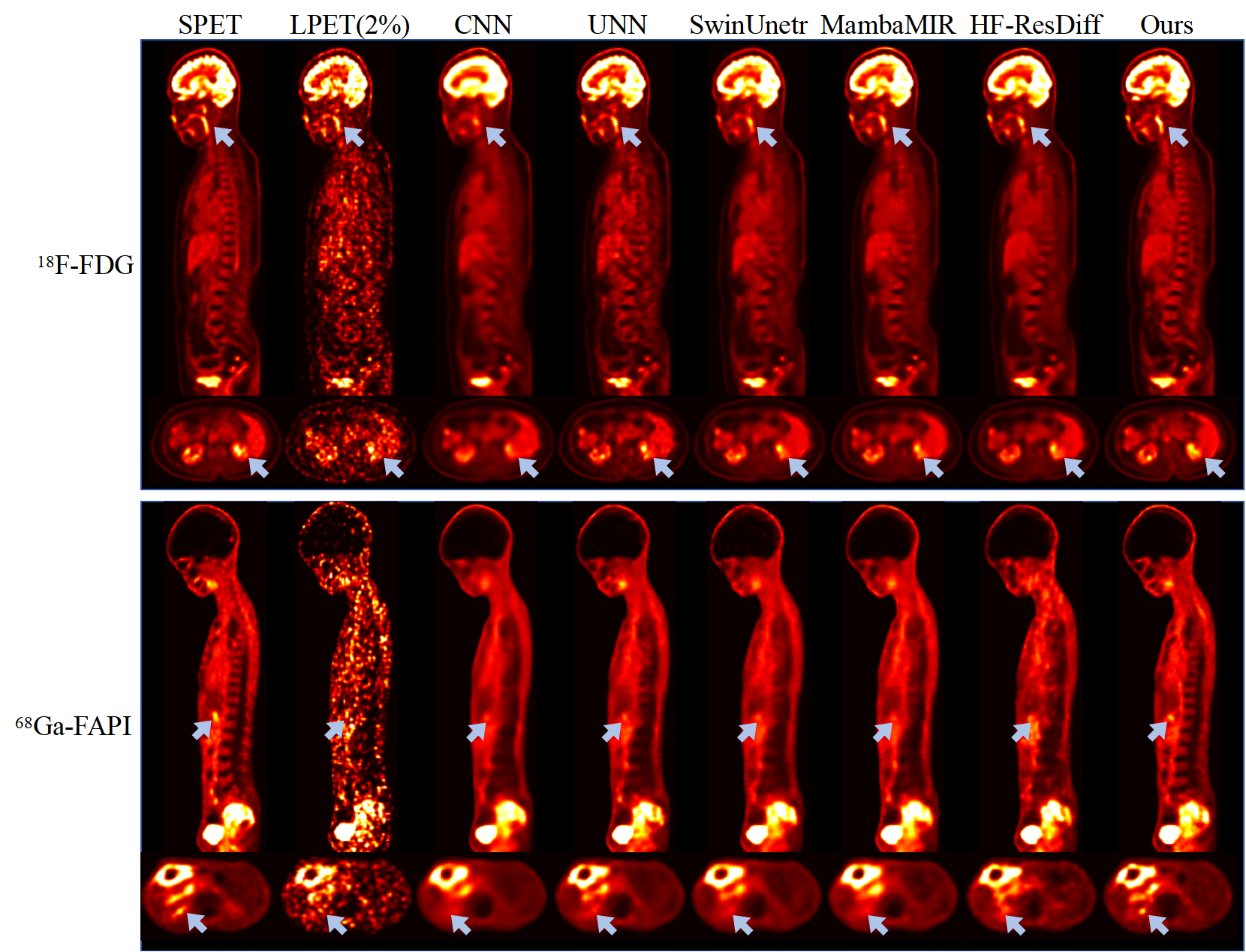}}
\caption{Quantitative comparison with state-of-the-art methods at 2\% dose level.}
\label{fig:compare}
\end{figure*}

\begin{table}[h!]
    \centering
    \caption{Quantitative ablation results on the $^{18}$F-FDG and $^{68}$Ga-FAPI tracer datasets.}
    \label{tab:ablation}
    \setlength{\tabcolsep}{8pt}
    \resizebox{\textwidth}{!}{
    \begin{tabular}{@{}lccccc@{}}
        \toprule
        \multicolumn{6}{c}{$^{18}$F-FDG} \\
        \cmidrule(lr){1-6}
        PSNR$\uparrow$/SSIM$\uparrow$ & 6s & 15s & 30s & 60s & 150s \\
        \midrule
        LPET           & 43.151/0.931 & 43.997/0.944 & 44.807/0.954 & 45.674/0.962 & 47.096/0.969 \\
        IDDPM          & 44.944/0.978 & 45.108/0.981 & 45.623/0.983 & 46.232/0.985 & 47.341/0.987 \\
        IDDPM+HWA      & 45.724/0.981 & 46.068/0.983 & 46.273/0.985 & 46.739/0.987 & 47.805/0.988 \\
        \textbf{Ours}  & \textbf{46.035/0.984} & \textbf{46.523/0.987} & \textbf{46.889/0.988} & \textbf{47.315/0.989} & \textbf{48.042/0.990} \\
        \midrule
        \multicolumn{6}{c}{$^{68}$Ga-FAPI} \\
        \cmidrule(lr){1-6}
        PSNR$\uparrow$/SSIM$\uparrow$ & 6s & 15s & 30s & 60s & 150s \\
        \midrule
        LPET           & 40.443/0.873 & 41.382/0.892 & 42.302/0.908 & 43.224/0.922 & 44.689/0.935 \\
        IDDPM          & 44.790/0.973 & 45.363/0.974 & 45.825/0.976 & 46.154/0.977 & 46.796/0.979 \\
        IDDPM+HWA      & 45.992/0.976 & 46.187/0.978 & 46.369/0.980 & 46.648/0.980 & 47.287/0.982 \\
        \textbf{Ours}  & \textbf{46.689/0.980} & \textbf{47.009/0.981} & \textbf{47.160/0.981} & \textbf{47.362/0.982} & \textbf{47.921/0.983} \\
        \bottomrule
    \end{tabular}
    }
\end{table}

\subsection{Ablation Study}
To validate the effectiveness of each proposed module, we conducted ablation studies with three sub-models: (1) IDDPM: the standard diffusion model; (2) IDDPM-HWA: IDDPM integrated with the CT-guided HWA; (3) MDAA-Diff (Ours): IDDPM incorporating both HWA and DAA module. All approaches were kept consistent in the experimental setup, ensuring a fair comparison (as shown in Table~\ref{tab:ablation}). The comparison results showed that IDDPM-HWA significantly improved denoising performance over IDDPM by leveraging CT fusion. MDAA-Diff further enhanced denoising capability through multi-dose adaptation and CT guidance, achieving clearer spinal and pulmonary boundaries with enhanced structural fidelity. These findings confirmed that the HWA and DAA module effectively improve PET denoising performance, validating their critical role in modeling multi-dose characteristics and anatomical constraints.

\section{Conclusion}
In this study, we propose a CT-Guided Multi-Dose Adaptive Attention Diffusion Model (MDAA-Diff) for LPET denoising. It utilizes high-frequency wavelet transforms from CT images to integrate multi-modal information. This approach address detail blurring limitations in traditional single-modal denoising models. Additionally, MDAA-Diff incorporates DAA module to explicitly model dose-dependent relationships, providing a robust and reliable solution for LPET imaging in clinical applications.
%
%
%
\bibliographystyle{splncs04}
\bibliography{ourbib1}
%

\end{document}